%% file: main.tex
\documentclass[10pt,sigconf]{acmart}

\usepackage{tikz}
\usetikzlibrary{shapes,shadows,arrows}

\tikzstyle{decision} = [diamond, draw]
\tikzstyle{line} = [draw, -stealth, thick]
\tikzstyle{elli}=[draw, ellipse, minimum height=8mm, text width=5em, text centered]
\tikzstyle{block} = [draw, rectangle,  text width=8em, text centered,node distance=7em]
\usetikzlibrary{positioning}
\usepackage{pifont}
\usepackage{float}
\usepackage{booktabs}
\floatstyle{plain}
\newfloat{example}{thp}{lop}
\floatname{example}{\textsc{Example}}
\usepackage{xcolor}
\usepackage[export]{adjustbox}
\usepackage{amsmath,amsfonts}
\usepackage{bm}
\usepackage{algorithmic}
\usepackage{efbox,graphicx}
\usepackage{textcomp}
\usepackage{xcolor}
\usepackage{pgfplots}
\usepackage{multirow}
\usepackage{subfigure}

\def\BibTeX{{\rm B\kern-.05em{\sc i\kern-.025em b}\kern-.08em
    T\kern-.1667em\lower.7ex\hbox{E}\kern-.125emX}}
    
\usepackage{amsthm}
\newcommand{\norm}[1]{\left\lVert#1\right\rVert}
 
\theoremstyle{definition}

\usetikzlibrary{matrix}

\usepackage[english]{babel}

\renewcommand\footnotetextcopyrightpermission[1]{} 
\setcopyright{none}

\acmDOI{}

\acmISBN{}  
\acmConference[Submitted for review to SIGCOMM]{}
\acmYear{2021}
\copyrightyear{}

\acmPrice{}
\begin{document}

\title{Traffic Flow Estimation using LTE Radio Frequency Counters and Machine Learning}


\author{Forough Yaghoubi}
\affiliation{%
   \institution{Ericsson AB}
   \city{Stockholm} 
   \state{Sweden} 
}
\email{forough.yaghoubi@ericsson.com}

\author{Armin Catovic}
\authornote{Previously employed at Ericsson AB}
\affiliation{%
   \institution{Schibsted Media Group}
   \city{Stockholm} 
   \state{Sweden} 
}
\email{armin.catovic@schibsted.com}

\author{Arthur Gusmao}
\affiliation{%
   \institution{Ericsson AB}
   \city{Stockholm} 
   \state{Sweden} 
}
\email{arthur.gusmao@ericsson.com}

\author{Jan Pieczkowski}
\affiliation{%
   \institution{Ericsson AB}
   \city{Stockholm} 
   \state{Sweden} 
}
\email{jan.pieczkowski@ericsson.com}

\author{Peter Boros}
\affiliation{%
   \institution{Ericsson AB}
   \city{Stockholm} 
   \state{Sweden} 
}
\email{peter.boros@ericsson.com}

\begin{abstract}
     As the demand for vehicles continues to outpace construction of new roads, it becomes imperative we implement strategies that improve utilization of existing transport infrastructure. Traffic sensors form a crucial part of many such strategies, giving us valuable insights into road utilization. However, due to cost and lead time associated with installation and maintenance of traffic sensors, municipalities and traffic authorities look toward cheaper and more scalable alternatives. Due to their ubiquitous nature and wide global deployment, cellular networks offer one such alternative. In this paper we present a novel method for traffic flow estimation using standardized LTE/4G radio frequency performance measurement counters. The problem is cast as a supervised regression task using both classical and deep learning methods. We further apply transfer learning to compensate that many locations lack traffic sensor data that could be used for training. We show that our approach benefits from applying transfer learning to generalize the solution not only in time but also in space (i.e., various parts of the city). The results are very promising and, unlike competing solutions, our approach utilizes aggregate LTE radio frequency counter data that is inherently privacy-preserving, readily available, and scales globally without any additional network impact.
\end{abstract}

\maketitle

\input{Introduction}
\input{Related}
\input{Approach}
\input{Transfer}

\input{Results}

\input{Ethical}
\input{Conclusion}

\section*{Acknowledgments}
We would like to thank the following people for their support throughout the project, and for facilitating the network and traffic sensor data without which none of this would be possible: Elin Allison, Madeleine Körling and Jyrki Lehtinen from Telia Company AB; Anders Broberg and Tobias Johansson from City of Stockholm; Annika Engström from KTH Royal Institute of Technology and Digital Demo Stockholm; Chris Deakin and Chris Holmes from WM5G Limited; Mo Elhabiby and Mike Grogan from Vodafone UK. We would also like to extend our gratitude to Leif Jonsson, Jesper Derehag, Carolyn Cartwright and Simone Ferlin, for reviewing our paper and providing valuable feedback.

\bibliographystyle{ACM-Reference-Format}
\bibliography{reference}

\end{document}

%% file: Introduction.tex
\section{Introduction}
The increasing number of vehicles in the public roadway network, relative to the limited construction of new roads, has caused recurring congestion in the U.S. and throughout the industrialized world \cite{tdhandbook2006}. In the U.S. alone, the total cost of lost productivity caused by traffic congestion was estimated at \$87 billion in 2018 \cite{fleming2019cost}. While one solution is to build new and expand existing roads, this is costly and takes time. A complementary approach is to implement strategies that improve the utilization of existing transport infrastructure. These strategies are found in Intelligent Transportation Systems (ITS) roadway and transit programs that have among their goals reducing travel time, easing delay and congestion, improving safety, and reducing pollutant emissions \cite{tdhandbook2006}.

Traffic flow sensor technology forms a key component of ITS. Traffic sensors can be categorized as in-roadway (e.g. inductive loop sensors and magnetometers), or over-roadway (e.g. traffic cameras, radar, infrared and laser sensors). More recently there has been a surge of ad-hoc over-roadway sensor technology, including road-side cellular network masts, Bluetooth and Wi-Fi sensors, as well as telemetry collected from connected vehicles, smartphones and GPS devices. Cellular network masts are particularly appealing, combining ubiquity of cellular network technology (e.g. LTE or more specifically E-UTRA), with strict high availability requirements.

While there have been previous approaches in utilizing cellular networks for understanding traffic flow, they've either been intrusive due to using user data, or not practical from a network operations perspective. In this paper we present a novel method for traffic flow estimation that leverages standard LTE/E-UTRA performance management (PM) counters, as defined by the 3rd Generation Partnership Project (3GPP) \cite{3gpp.32.425}. Namely, we utilize two radio frequency (RF) measurements - path loss distribution, and timing advance distribution counters, aggregated over 15 minute intervals. These counters are inherently privacy-preserving, and are continuously collected by nearly all LTE networks around the world, independent of network vendor. Thus our solution is non-invasive, and highly practical as it can be scaled across vast geographic regions with no live network impact.

Our contributions in this paper are threefold:
\begin{enumerate}
    \item We present a novel method for estimating traffic flow using classical and deep learning regression models trained on E-UTRA RF counters (features) and vehicle counts from actual traffic sensors (targets).
    \item We evaluate the performance of our models by applying the learned model to different time samples, referred to as temporal generalization in this paper.
    \item We evaluate the performance of our models by applying the learned model to different road segments lacking ground truth data, referred to as spatial generalization in this paper; it is shown that due to difference in traffic distribution, the performance of our models is sub-optimal, hence we improve the accuracy using two transfer learning approaches.
\end{enumerate}

The rest of the paper is organized as follows: in section \ref{section: related works} we summarize the related works in the area of traffic flow estimation; our overall solution including feature selection/transformation and learning algorithms is explained in section \ref{section: approach}; section \ref{section: transfer learning} introduces two different transfer learning approaches; in section \ref{section: results} we evaluate the performance of our models in terms of temporal and spatial generalization; ethical aspects are considered in section \ref{section: ethical considerations}; finally the key takeaways are summarized in section \ref{section: conclusion}.

%% file: Related.tex
\section{Related Works}~\label{section: related works}

Traffic flow estimation has and continues to be a popular research topic. In~\cite{zewei2015vehicle,george2013vehicle,ma2013wireless,haferkamp2017radio,nam2020deep} traffic flow estimation approaches are presented using data gathered from different sources such as cameras, acoustic sensors, magnetometers and spatially separated magnetic sensors. These solutions are not efficient due to coverage limitations and effort required in terms of installation and maintenance. To cope with these problems \cite{hansapalangkul2007detection,pattara2007estimating,hongsakham2008estimating,caceres2012traffic,xing2019traffic,ji2019deep,wang2020estimating} propose the use of mobile subscriber data in traffic flow estimation. In \cite{hansapalangkul2007detection,pattara2007estimating,hongsakham2008estimating} the cell dwelling time and global positioning system (GPS) coordinates of a mobile subscriber are used to estimate the traffic congestion. These methods however have the disadvantage of high power consumption on a mobile device due to constant use of GPS, and are inherently intrusive. Another type of cellular data is considered in \cite{caceres2012traffic} where authors propose a traffic flow estimation algorithm based on the number of subscribers in cars making a voice call. With today's heavy usage of streaming and social media services however, voice calls are hardly representative of the traffic density, which limits the accuracy of such an approach. The authors in \cite{xing2019traffic} use the travel trajectory of different mobile subscribers to detect in-vehicle users and henceforth compute the number of vehicles on a specific road. Tracking individual mobile users however is highly contentious and in most countries any user-identifiable or user-sensitive information limits the real-time usage of such data. In more recent work \cite{ji2019deep}, the authors propose a method to predict the traffic speed and direction using wireless communication access logs including S1 application protocol (S1AP) collected from multiple radio base stations (RBS) located within a predetermined distance from the road. Due to high-intensity nature of S1AP signalling, tracing on the S1 interface in every single RBS leads to increase in network load, which is undesirable as it could lead to network overload with potentially catastrophic consequences. Furthermore, S1AP exposes potentially sensitive subscriber information allowing for the user to be fingerprinted or tracked. Another approach as presented in \cite{wang2020estimating}, describes a "data fusion" approach, i.e. combining taxi GPS data with vehicle counts from license plate recognition (LPR) devices. The scalability of such approach is constrained due to limited availability of LPR and taxi GPS data.

Therefore in this paper, we propose a new solution to traffic flow estimation problem based on aggregate LTE/E-UTRA radio frequency counter data that is inherently privacy preserving, readily available, and does not impose any extra load on the network.

%% file: Approach.tex
\section{Method}\label{section: approach}
In this paper we describe two different approaches to using E-UTRA RF counters for traffic flow estimation. One approach involves using uplink path loss distribution as a feature vector in our model. Here we reason that different number of vehicles, i.e. obstacles on the road, can be represented by different path loss distributions. By optimizing for the number of vehicles the model should be able to discriminate between vehicles and all other users in the vicinity.

In the second approach we use radio propagation delay, or more specifically timing advance (TA), as our feature vector. LTE radio base stations (eNBs) estimate the propagation delay on every random access (RA) initiated by a user. These propagation delays are aggregated from all RAs and represented as a distribution over discretized distances, where every bin represents a certain distance range from the eNB. By selecting only the bins corresponding to the known distances between the eNB and the relevant road segments, we can directly capture the road users, i.e. vehicles.
\begin{figure}
    \begin{center}
        \includegraphics[width=0.51\textwidth]{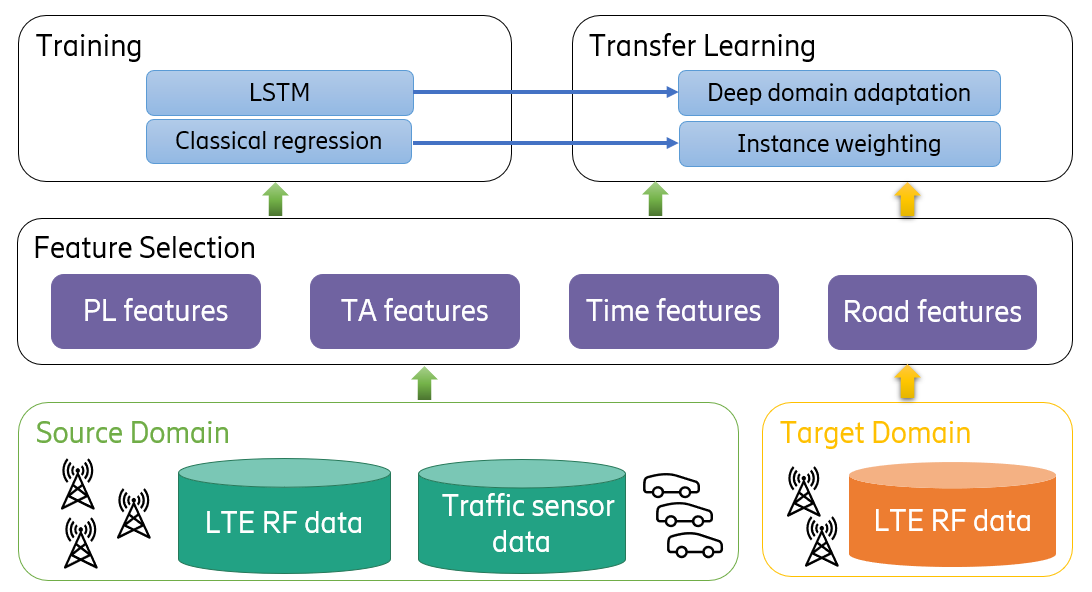}
    \end{center}
    \caption{Traffic flow estimation system presented in this paper.}
    \label{fig:system_overview}
\end{figure}

Path loss and timing advance features are described in detail in section \ref{sec: feature selection}. In both cases we use supervised regression techniques to train and evaluate our models. Fig. \ref{fig:system_overview} shows the high level view of our system. In this paper we work in two domains: the source domain consists of training and validation data - the feature and target variables; the second domain, referred to as target domain, is where we perform inference using only the feature variables - however we use ground truth data for evaluation purposes.

Our solution depends on the following assumptions:

\begin{itemize}
    \item The eNB, or more specifically the sector antenna, is located within line-of-sight (LOS) of the relevant road segment,
    \item The distance between the road segment and the sector antenna is known,
    \item The relevant road segment consists of predominantly vehicular traffic,
    \item Traffic sensors used to supply ground truth data completely capture the traffic flow along the relevant road segment.
\end{itemize}

In the following subsections we describe our feature and target variables, and learning algorithms.


\subsection{Feature and Target Variables}\label{sec: feature selection}

\begin{figure}
    \begin{center}
        \includegraphics[width=0.40\textwidth]{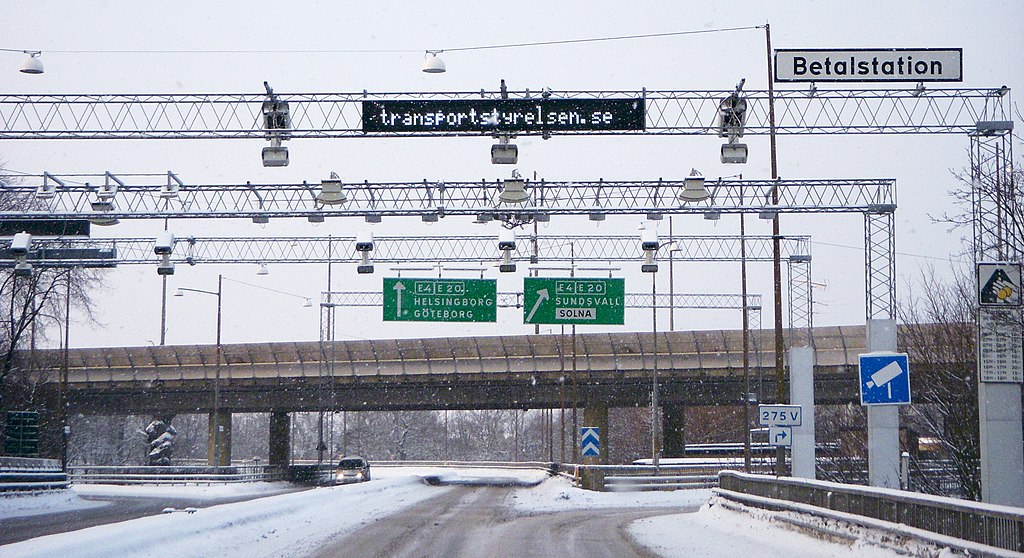}
    \end{center}
    \caption{An example of a laser-based traffic sensor used in this paper; photograph by Holger Ellgaard, distributed under a CC BY-SA 3.0 license.}
    \label{fig:traffic_sensor}
\end{figure}

\subsubsection{Traffic Sensor Data}
Target variables, i.e. ground truth data, consist of total vehicle counts aggregated over 15-minute intervals. The data is collected from a number of laser-based traffic sensors around inner Stockholm. Fig. \ref{fig:traffic_sensor} shows an example of such a sensor. Each sensor scans one lane of the road. For each road segment we sum the values from each lane to obtain total vehicle counts per 15-minute interval. We remove any samples where one of the sensor's values are missing (e.g. due to a malfunction).

\subsubsection{Path Loss Features}
Path loss (PL) is the attenuation of electromagnetic wave caused by free-space losses, absorption (e.g. by atmospheric particles), and scattering off various obstacles and surfaces. Radio propagation models attempt to account for this attenuation, and are a pivotal component in cellular network planning. Hata-Okamura \cite{hata1980propagation} models are one such family of radio propagation models used to approximate cellular network coverage in different environments. In LTE, eNBs estimate PL values for all the scheduled users on every transmit time interval (TTI), which is typically 1ms. These estimates, represented as decibel (dB) values are then placed into discretized PL bins; in our case we have 21 bins, where each bin covers a range of 5dB, starting from $<$50dB and going up to $>$140dB. These estimates are done per-frequency band; in our case we have three bands, 800MHz, 1800MHz (two separate antennas working in this band) and 2600MHz, so we concatenate PL bins for all bands, resulting in total of 4 x 21 = 48 PL features. We don't apply any filtering or transformation to PL features, and we treat all PL bins equally. Even though PL estimates are done every 1ms, the actual data available to us is aggregated in 15-minute intervals.

The motivation behind our use of PL features is that different traffic conditions will result in different radio wave scattering characteristics, leading to different path loss distributions. A condition where there are no vehicles on the road will be represented by path loss distribution $PL_{a}$, which would be representative of radio wave losses due to predominantly indoor users and pedestrians. On the other hand a condition where the traffic flow is greater than zero would result in path loss distribution $PL_{b}$ where $PL_{b} \neq PL_{a}$, since radio wave scattering off vehicle surfaces would yield a different path loss "signature". Our trained algorithms should be able to discriminate between such conditions.

\begin{figure}
    \begin{center}
        \includegraphics[width=0.40\textwidth]{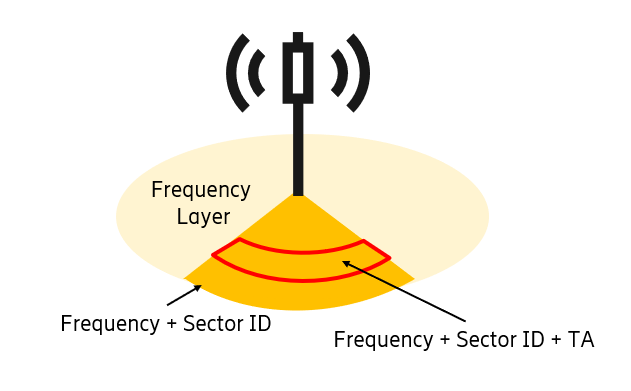}
    \end{center}
    \caption{Spatial granularity of an eNB.}
    \label{fig:spatial_granularity}
\end{figure}

\subsubsection{Timing Advance Features}
Timing advance (TA) is estimated for every user connection request, or more specifically on every random access. TA estimation is dependent on successful completion of an RRC Connection Request procedure, and the 11-bit TA command. The name is a slight misnomer, since TA features are actually represented as discretized distance bins/ranges, representing the distance between the user and the sector antenna. In our case we have 35 bins, starting from $<$80m up to 100km; typically only the first few bins are incremented, as users are normally within 500m of the antenna (otherwise they will be handed over to another sector, or another eNB). Fig. \ref{fig:spatial_granularity} shows the spatial granularity of an eNB and how the TA features may be represented.

Unlike PL features, we do actually apply a distance selection filter to TA features. Our aim is to consider only road users (vehicles), which means selecting TA bins/features that represent the known distances between the relevant road segment and the sector antenna. Lets assume that TA value ranges are indicated by $M$ bins where bin $ta_{i}$ corresponds to a distance interval shown by $[d_i; d_{i+1})$ given a known distance $d$ between the road segment and the sector antenna, we choose the TA bin index $ta_{i}$ where $d_{i} \leq d \leq d_{i+1}$.

Just like traffic sensor data and PL features, TA features are also aggregated in 15-minute intervals.

\subsubsection{Cyclic Time Features}
As traffic exhibits strong seasonality, it is beneficial to give our models temporal information. To encode this information, a common method is to transform the date-time representation into cyclic time features using a $\sin$ and $\cos$ transformation as follows:

\begin{align}
\nonumber
    x_{\sin}= & \sin(\frac{2\pi x}{\max(x)})\\ \nonumber
    x_{\cos}= & \cos(\frac{2 \pi x}{\max(x)})
\end{align}
where $x$ can be hour, day and month. By using the above equation, we convert time-of-day, day-of-week and week-of-month to the corresponding cyclic time features. As the time granularity for our data is in minutes, we set the $\max(x)$ to $24*60$, $7*24*60$, and $4*7*24*60$ respectively.

\subsubsection{Road-dependent Features}
As traffic flow depends on the road characteristics, we also apply different road-dependent features. These features are easily extracted from e.g. OpenStreetMap services. In this paper we use the following road-dependent features: number of lanes, maximum speed limit, and road category, i.e. highway, large city road and small city road.



\subsection{Learning Algorithms}
In this paper we compare two supervised learning approaches for traffic flow estimation. In the first approach we evaluate a number of different classical regression algorithms. In the second approach we take into account the history of time samples using gradient based Long Short-term Memory (LSTM).

\subsubsection{Classical Regression Models}
Classical regression assumes independence between time samples. Since we're working with fairly coarse 15-min aggregate intervals, it is reasonable to assume this independence. Regression then amounts to estimating a function $f(\bm{x}; \bm{\theta})$, which transforms a feature vector $\bm{x}$ to a target variable $y$. Function parameters $\bm{\theta} = (\theta_0, \theta_1, ..., \theta_k)$ are found by minimizing expected loss, typically a mean squared error (MSE) of the form $MSE(\bm{\theta}) = \frac{1}{N}\sum_{n=1}^N(y_n-f(\bm{x_n};\bm{\theta}))^2$, where $N$ corresponds to total number of 15-min aggregate samples. We evaluate a number of different regression algorithms including Support Vector Regressor (SVR), Kernel Ridge (KR), Decision Tree (DT), and Random Forest (RF). Each algorithm also requires setting its internal parameters, or hyperparameters. Since the total number of hyperparameters is small, we use grid search method to exhaustively search through the hyperparameter space and pick the combination of parameters that yield the best performance. We apply a time-dependent train/test split, e.g. by selecting the first 6 weeks for training, and the following 2 weeks for testing; compared to a random assignment of train/test data, our approach is more in line with how the algorithm would be used in practice, and is more representative of the generalization capability in the real-world setting.





\subsubsection{LSTM}


LSTM is a specific kind of recurrent neural network (RNN) that has the ability to capture long-term time dependencies and bridge time intervals in excess of 1000 steps even in case of noisy, in-compressible input sequences~\cite{du2017traffic}. Similar to other types of RNNs, LSTM has a chain structure with modified repeating modules. In each module, instead of having a single neural network layer, there are four layers that interact with each other. More detailed information about LSTM architecture can be found in~\cite{smagulova2019survey}. 




    
The architecture of our LSTM based traffic flow estimator consists of two LSTM layers, followed by a dropout regularization layer, and then finally the two fully-connected (FC) layers. The two LSTM, as well as the two FC layers, use the rectified linear unit (ReLU) activation function, while the output layer activates with the linear function.

%% file: Transfer.tex
\section{Transfer Learning Approaches}~\label{section: transfer learning}
The learning approaches mentioned above optimize the model for temporal generalization where we use all available locations in our training set, but withhold a contiguous period of time (e.g. two weeks) for test purposes. However, we would like our models to generalize well across all possible locations, even never-before-seen locations, which may potentially have completely different traffic patterns/distributions. We refer to this problem as spatial generalization. To cope with this problem we use transfer learning (TL) approaches. TL focuses on transferring the knowledge between different domains and can be a promising solution to overcome the spatial generalization problem. Recently, there has been lot of work focusing on transfer learning and proposing efficient solutions~\cite{zhuang2019comprehensive,pan2009survey,tan2018survey}. These studies categorize TL into three subcategories based on different situations involving source and target domain data and the tasks, including inductive, transductive, and unsupervised transfer learning. Our work can be fitted into transductive transfer learning where the source label data are available while no label data for target domain is provided. Here the assumption is that the task between target and source domain is the same, but the domain marginal or conditional distributions are different. Among the proposed transductive TL algorithms, we evaluate two approaches - one based on instant weighting and the second one based on deep domain adaptation. We explain each of the algorithms in detail in the following sections.

\subsection{Instant Weighting}
The data-based TL approaches, such as instant weighting, focus on transferring the knowledge by adjustment of the source data. Assuming that the source and target domain only differ in marginal distribution, a simple idea for transformation is to assign weights to source domain data equal to the ratio of source and target domain marginal distribution. Therefore the general loss function of the learning algorithm is given by:
\begin{equation}
    \min_{\theta} \frac{1}{N_{s}}\sum_{1}^{N_{s}}\alpha_{i}J(\theta(x_{i}^{s}),y_{i}^{s}) + \lambda \gamma(\theta)
\end{equation}
where $J$ represents the loss of source data and $\alpha_{i}$ is the weighting parameter and is equal to:
\begin{equation}
    \alpha_{i}=\frac{P^{T}(x)}{P^{S}(x)}.
\end{equation}

In the literature, there exist many ways to compute $\alpha_{i}$; in~\cite{huang2006correcting} the authors used Kernel Mean Matching (KMM) to estimate the ratio by matching the means of target and source domain data in the reproducing-kernel Hilbert space where the problem of finding weights can be written as follows:

\begin{align}
    \min_{\alpha} \frac{1}{2} \alpha^{T} K \alpha - \kappa \alpha\\
    \nonumber s.t \sum_{i}^{N_s} \alpha_{i} - N_{s} \leq N_{s} \epsilon\\
    \nonumber \alpha \in [0, B]
\end{align}


where $N_{s}$ shows the number of sample in source domain data and $K$ is kernel matrix and is defined as:
\begin{equation}
K=\begin{bmatrix}
k_{ss} & k_{st} \\ 
k_{ts} & k_{tt} 
\end{bmatrix},    
\end{equation}
while $k_{ss}=k(x_{s},x_{s})$ and $\kappa_{i}=\frac{N_{s}}{N_{T}}\sum_{i}^{N_{T}}k(x_{i},x_{Tj})$.

\subsection{Domain Adaptation}
Deep learning algorithms have received lot of attention from researchers having successfully outperformed many traditional machine learning methods in tasks such as computer vision and natural language processing (NLP). Therefore in the TL area many researchers also utilize deep learning techniques.

In this paper, we use discrepancy-based domain adaptation, where a deep neural network is used to learn the domain-independent feature representations. In deep neural networks, the early layers tends to learn more generic transferable features, while domain-dependent features are extracted in the terminal layers. Therefore, to decrease the gap between the distribution in the last layers, we add multiple adaptation layers with discrepancy loss as regularizer. 

\begin{figure}
    \begin{center}
        \includegraphics[width=0.48\textwidth]{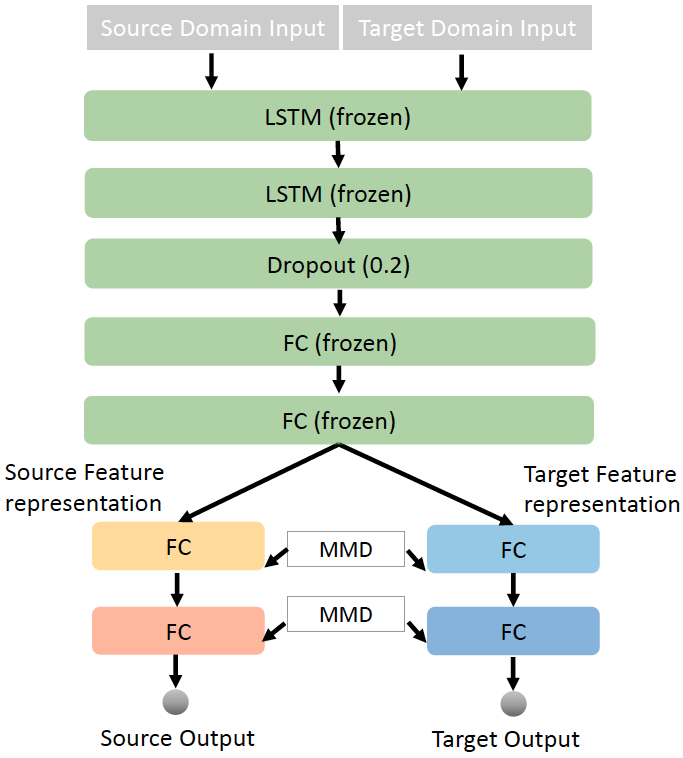}
    \end{center}
    \caption{The domain adaptation network based on LSTM.}
    \label{fig: domain adapatation arch}
\end{figure}

The deep learning model used for feature extraction is the LSTM model explained in previous section. The pretrained LSTM model will be used to extract the features for both source and target domains. After that the primary goal is to reduce the difference between target and source domain distribution. The term maximum mean discrepancy (MMD) is widely used in TL literature as a metric to compute the distance between two distribution~\cite{wang2017vehicle,gretton2012optimal}. Fig. \ref{fig: domain adapatation arch} shows the architecture of our domain adaptation network based on LSTM.

Let $f$ denote the function for feature representation of our pretrained model, then the distance between the feature distribution of source and target domain is given by:
\begin{equation}\label{eq: mmd}
d(p,q)=\sup_{f\in F}{E_{p}\{f(x)\} -E_{q}\{f(y)\}}  
\end{equation}
where $\sup$ defines the supremum, $E$ denotes the expectation and $x$ and $y$ are independently and identically distributed (i.i.d) samples from $p$ and $q$, respectively. The above equation can be easily computed using the kernel trick where it can be expressed by expectation of kernel functions. Therefore, the square of equation~\eqref{eq: mmd} can be reformulated as follows:
\begin{equation}
d^{2}_{k}(p,q)=E_{x_{p}^{s}x_{p}^{s}k(x_{p}^{s},x_{p}^{s})} + E_{x_{q}^{t}x_{q}^{t}k(x_{q}^{t},x_{q}^{t})}  - 2 E_{x_{p}^{s}x_{q}^{t}k(x_{p}^{s},x_{q}^{t})},    
\end{equation}
where $x_{p}^{s}$ and $x_{q}^{t}$ are the samples from source and target domain respectively, and $k$ is the kernel defined as $\exp(\frac{-\norm{x_{i}-x_{j}}^{2}}{\gamma})$. 

To adapt the pretrained model for the target data samples, the objective function of our TL algorithm is given by~\cite{long2015learning}:

\begin{equation}
    \min_{\theta} \frac{1}{N_{s}} \sum_{1}^{N_{s}} J(\theta(x_{i}^{s}),y_{i}^{s}) + \lambda \sum_{l=l_{1}}^{l_{2}} d_{k}^{2}(D^{s}_{l},D^{t}_{l}),
\end{equation}
where $J$ is the loss for source domain in LSTM network, $l_{1}$ and $l_{2}$ indicate the layer indices between which the regularization is effective, and $D^{s}_{l}$ and $D^{t}_{l}$ are $l$ layer representation of the source and target samples, respectively. The parameter $\lambda$ is a trade off term so that the objective function can benefit both from TL and deep learning.

%% file: Results.tex
\section{Results}\label{section: results}
We use approximately 8 weeks worth of data, where every data sample corresponds to a 15-min interval, so we have {\raise.17ex\hbox{$\scriptstyle\sim$}} 96 * 7 * 8 = 5376 data samples. The data are collected from six different locations around inner Stockholm; each location corresponds to a road segment with a traffic sensor and a nearby LTE eNB. We evaluate models using PL and TA features independently and across a range of regression algorithms. When evaluating temporal generalization we use all locations during training and split the data into 80/20 train/test sets, which corresponds to approximately 6 weeks of contiguous training data, and 2 weeks of test data. When evaluating spatial generalization we use all time samples for training but we randomly assign road segments into source and target domains. For evaluation purposes we use coefficient of determination $R^2$ defined as follows:

\begin{equation}
    R^{2}=1-\frac{SS_{res}}{SS_{tot}}
\end{equation}
where 
\begin{align} \nonumber
    & SS_{tot}=\sum_{i}\left(y_{i}-\bar{y}\right)^2 \\
    & SS_{res}=\sum_{i} \left(y_{i}-\hat{y}_{i}\right)^2
\end{align}
$SS_{tot}$ represents total sum of squares, and $SS_{res}$ represents residual sum of squares, while $\bar{y}$ and $\hat{y_i}$ are the observed data mean and the predicted traffic flow respectively. A model that always predicts observed data mean will have $R^2=0$; models with observations worse than the observed data mean will have negative values; the most optimal value is $R^2=1$, so we want our models to be as close to 1 as possible.

The set of classical regression algorithms used for training are Support Vector Regression (SVR), Kernel Ridge (KR), Decision Trees (DT) and Random Forest (RF). We also train a deep learning model with two LSTM layers followed by a dropout layer and two fully-connected layers activated with the ReLU function. The hyperparameters providing the best $R^2$ score on the test set for our models are found using grid search and presented in Table~\ref{table: hyper_parameter_telia}. The corresponding results for both temporal and spatial generalization performance are shown in Table~\ref{table: r2_score}.

\begin{table}
\centering
\caption{Hyperparameters used in this paper.}\label{table: hyper_parameter_telia}
\renewcommand{\tabcolsep}{20pt}
\renewcommand{\arraystretch}{1.3}
    \begin{tabular}{c | c}
     \hline \hline
     Models & Parameters \\ \hline
        \multirow{3}{*} & Kernel = rbf    \\ 
            SVR       & $C$ = 10      \\
                       & $\gamma$ = 0.001 \\ \hline
        \multirow{3}{*} & Kernel = rbf    \\ 
            KR       & $\alpha$ = 1      \\
                       & $\gamma$ = 0.01 \\ \hline
     DT & Maximum depth = 10  \\ \hline
     RF & Maximum depth = 30  \\ \hline
     \multirow{5}{*} & Learning rate = 0.0009   \\ 
      LSTM             & Hidden size = 100      \\
                       & Epochs = 300 \\
                       & Dropout rate = 0.2 \\
                    & Window = 5 \\ \hline \hline 
    \end{tabular}
\end{table}

\begin{table}
\centering
\caption{$R^2$ score for temporal and spatial generalization performance using TA and PL features. Higher scores are better.}\label{table: r2_score}
\renewcommand{\tabcolsep}{7.5pt}
\renewcommand{\arraystretch}{1.3}
\begin{tabular}{c|c|c|c|c}
\hline\hline
\multirow{2}{*}{Models} & \multicolumn{2}{l|}{Temporal Generalization} & \multicolumn{2}{l}{Spatial Generalization} \\ \cline{2-5} 
                        & TA        & PL       & TA        & PL        \\ \hline
SVR                     & 0.754             & 0.786            & 0.12              & -0.62             \\ \hline
KR                      & 0.862             & 0.888            & -0.79             & -0.63             \\ \hline
DT           & 0.938             & 0.946            & -3.22             & -0.37             \\ \hline
\textbf{RF}           & \textbf{0.946}             & \textbf{0.959}            & -0.96             & 0.017             \\ \hline
LSTM                    & 0.845             & 0.901            & 0.087             & -1.67             \\ \hline
\end{tabular}
\end{table}

\begin{figure}
	\hfill
	\subfigure[\label{fig: road_1}]{\includegraphics[width=0.98\columnwidth]{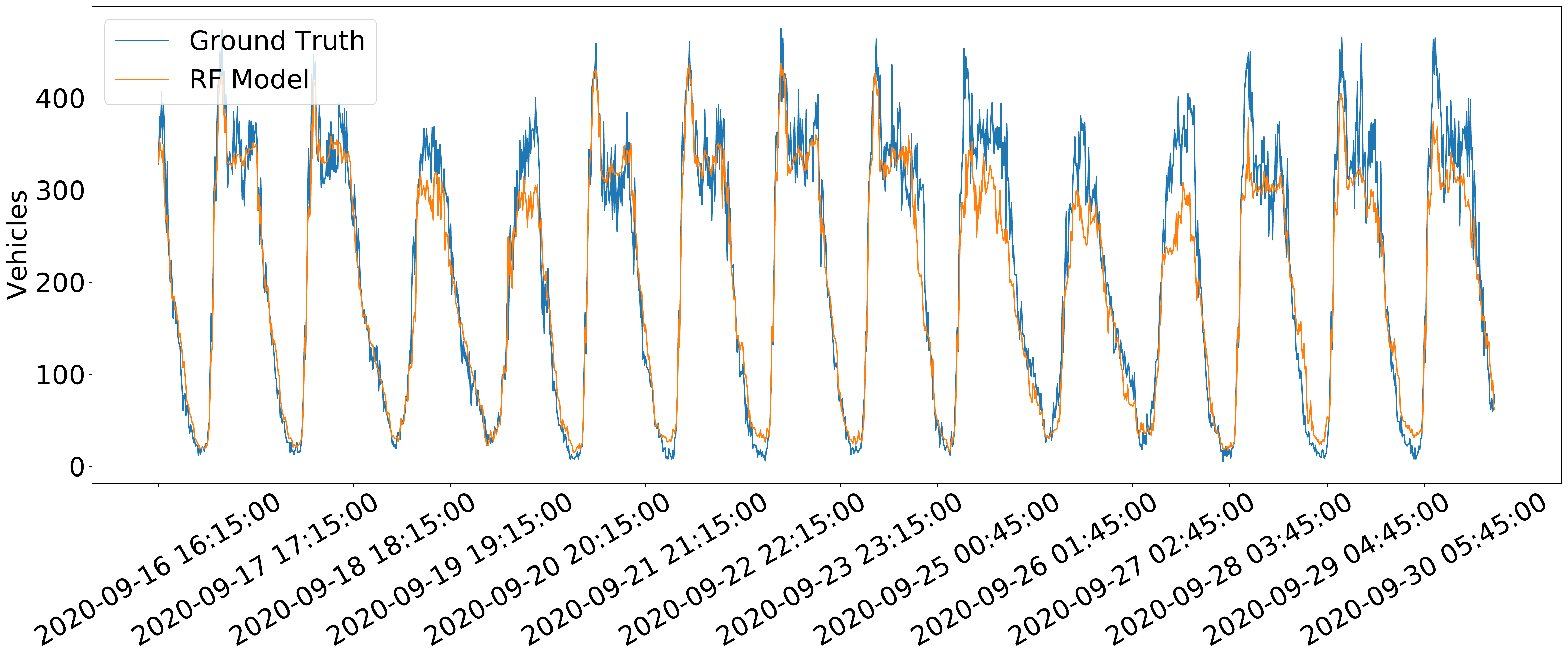}}
	\hfill
	\subfigure[\label{fig: road_2}]{\includegraphics[width=0.98\columnwidth]{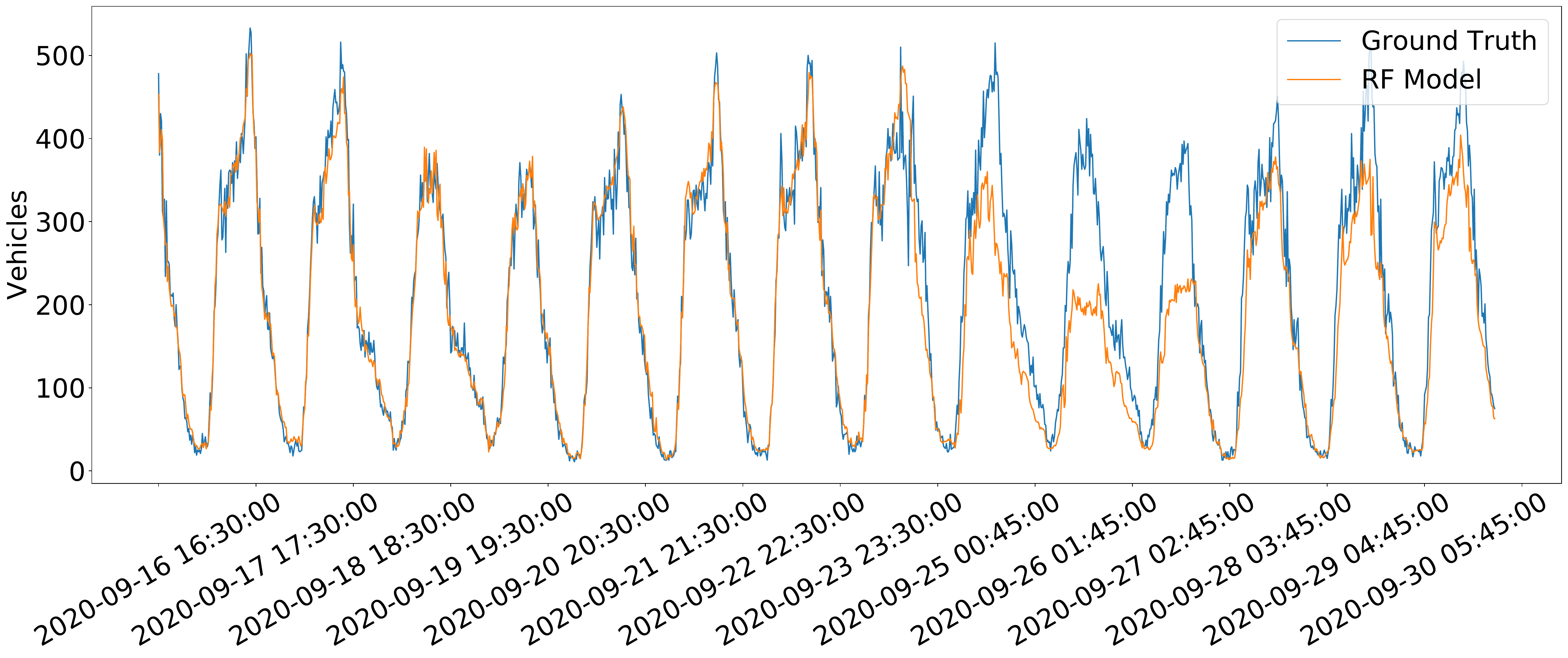}}
	\hfill
	\subfigure[\label{fig: road_3}]{\includegraphics[width=0.98\columnwidth]{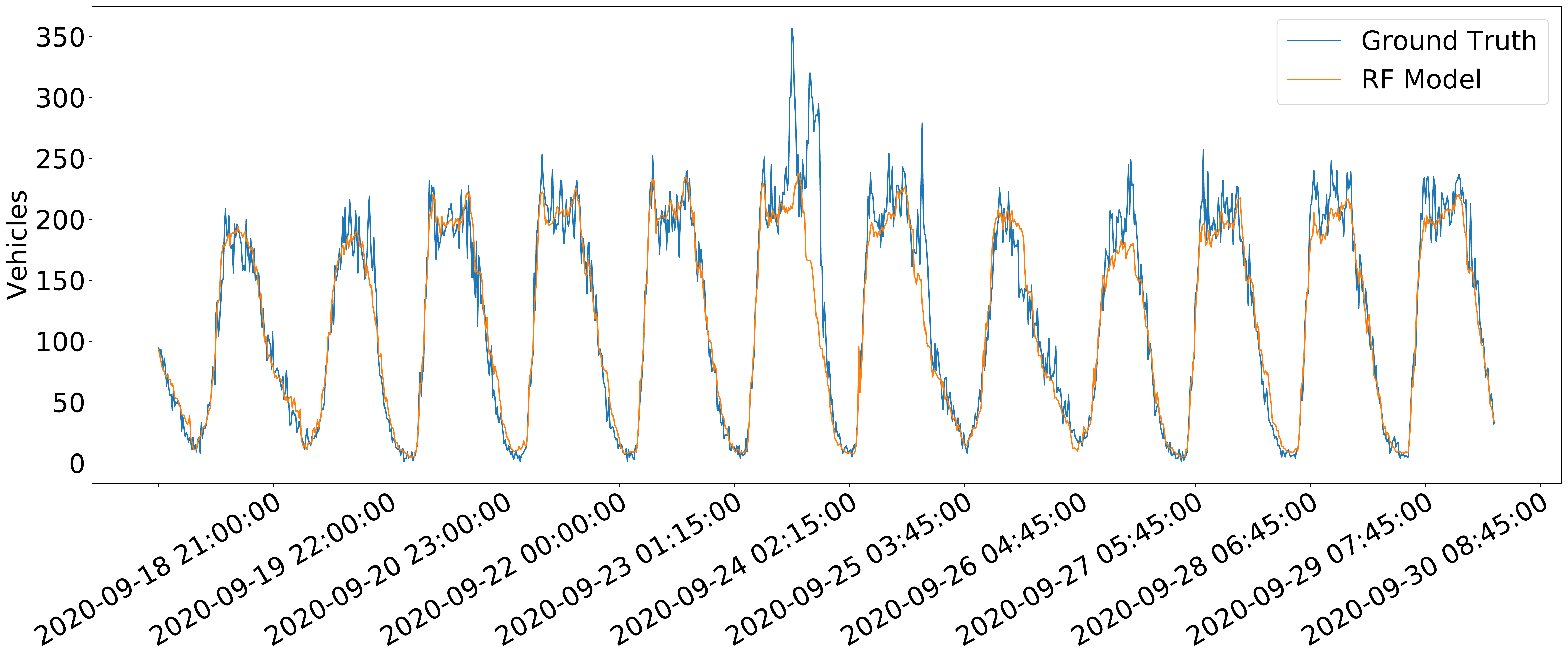}}
	\hfill
	\caption{Traffic flow estimates in terms of number of vehicles per 15-min interval using our Random Forest (RF) model compared to the ground truth, for \subref{fig: road_1} Road 1, \subref{fig: road_2} Road 2 and \subref{fig: road_3} Road 3.}
	\label{fig: time_generlization}
\end{figure}

The results in Table \ref{table: r2_score} indicate that all regression algorithms perform reasonably well in terms of temporal generalization, using either TA or PL features. The Random Forest (RF) model outperforms all the others, including the LSTM model, with an average $R^2$ score of 0.95. These results validate our initial assumption that due to a fairly coarse 15-min aggregate interval, it is safe to assume independence between time steps, hence deep learning based LSTM does not add any additional value. A more visual representation of the RF algorithm performance is shown in Fig. \ref{fig: time_generlization} where we compare traffic flow estimates from our model against the actual values across three different locations. The algorithm does not always capture the peaks - our hypothesis is that more training samples with varied traffic flow distributions are needed for the model to generalize even better.

Despite good temporal generalization performance, the average $R^2$ score for spatial generalization is very low for all regression models. This poor performance is due to inherent difference between the source and target domain distributions. In order to improve spatial generalization we use two types of transfer learning (TL) algorithms, namely instant weighting and deep domain adaptation.

In the first approach we implement the instant weighting for classical regression. For each test location, we compute the weights solving the quadratic optimization problem, and then retrain the model using these weights. Since the RF model yields the highest $R^2$ score on temporal generalization we apply instant weighting to RF only. 

Table \ref{table: instant-weighting r2} presents the $R^2$ scores of RF model for both TA and PL features with and without applying the instant weighting. The results indicate that instant weighting can only improve the performance when TA features are used. Since the TA features represent the road users more explicitly we expect there to be some minimum similarity between all domain distributions. On the other hand PL represents all users, including indoor users, and therefore PL features are highly sensitive to physical layout of the environment, i.e. number of buildings, thickness of walls, heights of buildings etc.

\begin{table}
\centering
\caption{$R^2$ score for spatial generalization using the Random Forest (RF) model with and without transfer learning (TL). Higher scores are better.}\label{table: instant-weighting r2}
\renewcommand{\tabcolsep}{12pt}
\renewcommand{\arraystretch}{1.3}
\begin{tabular}{c|c|c|c|c}
\hline\hline
\multirow{2}{*}{Test Road} & \multicolumn{2}{l|}{PL Features} & \multicolumn{2}{l}{TA Features} \\ \cline{2-5} 
                      & No TL            & TL            & No TL        & TL            \\ \hline
1                     &   0.02         &   -0.47       &    -0.96          &  \textbf{0.71}            \\ \hline
2                     &   0.02          &   -0.75      &     -0.96        &    \textbf{0.72}           \\ \hline
3                     &    0.02         &   -0.86      &      -0.96        &     \textbf{0.42}          \\ \hline\hline
Mean                     &   0.02        &   -0.69      &      -0.96        &     \textbf{0.62}         \\ \hline

\end{tabular}
\end{table}

\begin{table}
\centering
\caption{$R^2$ score for spatial generalization using the LSTM model with and without transfer learning (TL). Higher scores are better.}\label{table: domain_adaptaion r2}
\renewcommand{\tabcolsep}{12pt}
\renewcommand{\arraystretch}{1.3}
\begin{tabular}{c|c|c|c|c}
\hline\hline
\multirow{2}{*}{Test Road} & \multicolumn{2}{l|}{PL Features} & \multicolumn{2}{l}{TA Features} \\ \cline{2-5} 
                      & No TL            & TL            & No TL        & TL            \\ \hline
1                     &   -0.20         &   0.24       &    0.24          &  \textbf{0.66}            \\ \hline
2                     &   -1.73          &   -1.02      &     -0.62        &    \textbf{0.61}           \\ \hline
3                     &    -3.09         &   -0.52      &      -0.13        &     \textbf{0.61}          \\ \hline\hline
Mean                     &   -1.67        &   -0.43      &      -0.17        &     \textbf{0.63}         \\ \hline

\end{tabular}
\end{table}

In the second approach, we implement the deep domain adaptation algorithm as shown in Fig. \ref{fig: domain adapatation arch}. We freeze the two LSTM layers and the two fully-connected layers using the pre-trained weights, while we train the final two fully-connected layers using the MMD regularizer. As there is no target domain label data available only the source output is considered in the loss function.

Table \ref{table: domain_adaptaion r2} shows the performance of spatial generalization using the LSTM and deep domain adaptation. The LSTM model performs reasonably well using TA features, with average $R^2$ score very similar to what we saw using RF and instant weighting.

%% file: Ethical.tex
\section{Ethical Considerations}\label{section: ethical considerations}
One of the main motivations for the work presented in this paper concerns user privacy and integrity. Traffic cameras and automated license plate recognition devices allow for unprecedented levels of identification and tracking. This is all the more true for user data obtained from cellular networks and mobile devices. Our approach as presented in this paper uses data that is inherently 
privacy-preserving - we use readily available radio frequency counters that are aggregated on cell level and per definition do not contain any information about individual users, nor could this information be reconstructed. It is therefore impossible to identify or track any individual user based on this data. With that in mind we can state that the work presented in this paper does not raise any ethical issues.

%% file: Conclusion.tex
\section{Conclusion}~\label{section: conclusion}
Traffic flow estimation has traditionally involved forecasting methods based on observations from dedicated traffic sensors. Firstly these methods don't scale well since we require large number of sensors. Secondly we would need a separate forecasting model for every road, since roads don't exhibit homogeneous behaviour. Finally our traffic estimation performance would be susceptible to drastic changes in driver behaviour or road conditions, such as traffic accidents and road works. To overcome these limitations alternative approaches have been proposed, including using various forms of cellular network data to estimate traffic flow. However existing approaches are either user invasive, or can potentially result in adverse operational impacts to cellular networks.

In this paper we propose a traffic flow estimation method using inherently anonymous and widely available LTE/E-UTRA radio frequency counters, namely path loss and timing advance counters, effectively turning LTE eNBs into traffic sensors. We cast traffic flow estimation as a supervised regression problem, where path loss and timing advance counters are used as primary features, and vehicle counts from actual traffic sensors as target or ground truth variables. We demonstrated excellent performance using both Random Forest and LSTM regression models. Since we have limited amount of ground truth data, i.e. we only had access to six different locations, we also evaluated the performance of two different transfer learning approaches, namely instant weighting, and deep domain adaptation. With transfer learning we demonstrated reasonable performance using either Random Forests or LSTMs, but using only timing advance features. Our hypothesis is that with more data and more locations the performance will improve further still.

While our models are not perfect estimators, they are still extremely useful - they capture the shape of the traffic very well, and for most purposes provide a good-enough estimate of the traffic flow. The output of these models can be used for anomaly detection, for example for detecting traffic congestion or accidents. All this can be achieved without having to install any additional sensors - we simply re-use LTE radio base stations that are permanently fixed in their locations with near 100\% uptime.